\documentclass[letter]{aa}

\usepackage[varg]{txfonts}

\usepackage{graphicx}
\usepackage{amsmath}

\usepackage[colorlinks=true,linkcolor=magenta,citecolor=blue,filecolor=black,urlcolor=cyan,final=true]{hyperref}

\makeatletter
\renewcommand*\aa@pageof{, page \thepage{} of \pageref*{LastPage}}
\makeatother

\begin{document}

   \title{Prominence eruption observed in He~II 304~{\AA} up to ${>}6\,R_\odot$ \\by EUI/FSI aboard Solar Orbiter}

   \author{
   M.~Mierla\inst{\ref{ROB},\ref{RomanianAcademy}}
   \and
   A.~N.~Zhukov\inst{\ref{ROB},\ref{SINP}}
   \and
   D.~Berghmans\inst{\ref{ROB}}
   \and
   S.~Parenti\inst{\ref{IAS}}
   \and 
   F.~Auch{\`e}re\inst{\ref{IAS}}
   \and
   P.~Heinzel\inst{\ref{CAS},\ref{UWroclaw}}
   \and
   D.~B.~Seaton\inst{\ref{SwRI}}
   \and
   E.~Palmerio\inst{\ref{PSI}}
   \and
   S.~Jej\v{c}i\v{c}\inst{\ref{Lubjana1},\ref{Lubjana2}}
   \and
   J.~Janssens\inst{\ref{ROB}}
   \and
   E.~Kraaikamp\inst{\ref{ROB}}
   \and
   B.~Nicula\inst{\ref{ROB}}
   \and
   D.~M.~Long\inst{\ref{MSSL}}
   \and
   L.~A.~Hayes\inst{\ref{ESA}}
   \and
   I.~C.~Jebaraj\inst{\ref{ROB}}
   \and
   D.-C.~Talpeanu\inst{\ref{ROB}}
   \and
   E.~D’Huys\inst{\ref{ROB}}
   \and
   L. Dolla\inst{\ref{ROB}}
   \and
   S.~Gissot\inst{\ref{ROB}}
   \and
   J.~Magdaleni{\'c}\inst{\ref{ROB}}
   \and
   L.~Rodriguez\inst{\ref{ROB}}
   \and
   S.~Shestov\inst{\ref{ROB}}
   \and
   K.~Stegen\inst{\ref{ROB}}
   \and
   C.~Verbeeck\inst{\ref{ROB}}
   \and
   C.~Sasso\inst{\ref{INAF-OAC}}
   \and
   M.~Romoli\inst{\ref{UniFi},\ref{INAF}}
   \and
   V.~Andretta\inst{\ref{INAF-OAC}}
     }

  \institute{
  Solar--Terrestrial Centre of Excellence --- SIDC, Royal Observatory of Belgium, 1180 Brussels, Belgium\\
  \email{marilena.mierla@oma.be}
  \label{ROB}
  \and
  Institute of Geodynamics of the Romanian Academy, 020032 Bucharest-37, Romania
  \label{RomanianAcademy}
  \and
  Skobeltsyn Institute of Nuclear Physics, Moscow State University, 119992 Moscow, Russia
  \label{SINP}
  \and
  Universit{\'e} Paris-Saclay, CNRS, Institut d'Astrophysique Spatiale, 91405 Orsay, France
  \label{IAS}
  \and
  Astronomical Institute of the Czech Academy of Sciences, 251 65 Ond\v{r}ejov, Czech Republic
  \label{CAS}
  \and
  University of Wroc\l{}aw, Center of Scientific  Excellence - Solar and Stellar Activity, Kopernika 11, 51-622 Wroc\l{}aw, Poland
  \label{UWroclaw}
   \and
  Southwest Research Institute, Boulder, CO 80302, USA
  \label{SwRI}
  \and
  Predictive Science Inc., San Diego, CA 92121, USA
  \label{PSI}
  \and
  Faculty of Education, University of Ljubljana, 1000 Ljubljana, Slovenia
  \label{Lubjana1}
  \and
  Faculty of Mathematics and Physics, University of Ljubljana, 1000 Ljubljana, Slovenia
  \label{Lubjana2}
  \and
  Mullard Space Science Laboratory, University College London, Holmbury St. Mary, Dorking, Surrey, RH5 6NT, UK
  \label{MSSL}
  \and
  ESTEC, European Space Agency, 2201 AZ Noordwijk, The Netherlands
  \label{ESA}
  \and
  INAF --- Osservatorio Astronomico di Capodimonte, I-80131 Naples, Italy
  \label{INAF-OAC}
  \and
  Dipartimento di Fisica e Astronomia, Universit{\`a} di Firenze, I-50019 Sesto Fiorentino FI, Italy
  \label{UniFi}
  \and
  INAF --- Associate Scientist
  \label{INAF}
             }

   \date{}

 
  \abstract
  {}
   {We report observations of a unique, large prominence eruption that was observed in the He~II 304~{\AA} passband of the the Extreme Ultraviolet Imager/Full Sun Imager telescope aboard Solar Orbiter on 15--16 February 2022.}
   {Observations from several vantage points -- Solar Orbiter, the Solar–Terrestrial Relations Observatory, the Solar and Heliospheric Observatory, and Earth-orbiting satellites -- were used to measure the kinematics of the erupting prominence and the associated coronal mass ejection. Three-dimensional reconstruction was used to calculate the deprojected positions and speeds of different parts of the prominence. Observations in several passbands allowed us to analyse the radiative properties of the erupting prominence. }
   {The leading parts of the erupting prominence and the leading edge of the corresponding coronal mass ejection propagate at speeds of around $1700$~km~s$^{-1}$ and $2200$~km~s$^{-1}$, respectively, while the trailing parts of the prominence are significantly slower (around 500~km~s$^{-1}$). Parts of the prominence are tracked up to heights of over $6\,R_{\odot}$. The He~II emission is probably produced via collisional excitation rather than scattering. Surprisingly, the brightness of a trailing feature increases with height.}
   {The reported prominence is the first  observed in He~II 304~{\AA} emission at such a great height (above 6~$R_\odot$).}


   \maketitle
%

\section{Introduction}
\label{S-intro}

Prominences (or filaments, when they are seen projected on the solar disk) are cool and dense plasma structures observed in the hot and tenuous solar corona \citep[see e.g.][]{Labrosse2010, Mackay2010, Bemporad2011, Parenti2014, Gibson2018}. When erupting, prominences are often associated with coronal mass ejections (CMEs), which can arrive at Earth and trigger geomagnetic storms. In quiescent prominences, one often detects dark absorbing features in various extreme ultraviolet (EUV) coronal passbands \citep[e.g.][]{Parenti2012}. Cool prominence plasma absorbs background coronal EUV line radiation via the photoionisation of hydrogen, neutral helium, and singly ionised helium. 
Another mechanism is `emissivity blocking', but this is mostly negligible in comparison to absorption \citep{Anzer2005}.
While in the 193~{\AA} passband one usually sees absorption, the 171~{\AA} and 131~{\AA} passbands also show portions of prominences in emission, which \citet{Parenti2012} attributed to the prominence--corona transition region (PCTR) at temperatures below 1\,MK. 

The evolution of an erupting prominence from its quasi-equilibrium state through eruption can be described by four phases: the quasi-static phase, the initiation phase (slow rise), the impulsive main acceleration phase (fast rise), and the propagation phase with only slowly varying velocity \citep[e.g.][]{Zhang2006, Liu2021}.
In order to trigger the initiation phase, different mechanisms are invoked \citep[see e.g.\ the review by][]{Green2018}, such as tether-cutting \citep[e.g.][]{vanBallegooijen1989, Moore2001}, flux cancellation \citep[e.g.][]{Amari2003, Linker2003}, magnetic breakout \citep[e.g.][]{Antiochos1999}, flux emergence \citep[e.g.][]{Chen2000}, and mass-loading \citep[e.g.][]{Seaton2011}, as well as the onset of ideal magnetohydrodynamic (MHD) instabilities, including kink \citep[e.g.][]{Torok2005} and torus instabilities \citep[e.g.][]{Kliem2006}, and loss of equilibrium \citep[e.g.][]{Priest2002}. 

The acceleration phase occurs when the erupting prominence experiences a sudden increase in velocity. This phase is driven either by magnetic reconnection taking place at the site of the underlying current sheet \citep[e.g.][]{Lin2000, Karpen2012} or by an ideal MHD instability \citep[e.g.][]{Fan2007, Demoulin2010}. 
After the acceleration phase, most erupting filaments rise with an almost-constant velocity \citep[e.g.][]{Kahler1988, Joshi2007}. This defines the propagation phase. 
Different kinematic characteristics indicate that different physical processes dominate in these phases. For example, Lorentz forces are dominant during the acceleration phase, while during the propagation phase the Lorentz force is comparable to the gravitational and drag forces \citep{Chen2003, Temmer2011, Majumdar2020}. 

The initiation and acceleration phases of prominence eruptions close to the Sun are usually observed by EUV telescopes, while the propagation phase of corresponding CMEs farther out is usually covered by white-light coronagraphs
\citep{Mierla2013, Reva2017}. 
Connecting low-corona EUV observations to high-corona white-light observations was attempted, for example, by \citet{Byrne2014}, \citet{Dhuys2017}, and \citet{OHara2019}.

Since the launch of the Solar Orbiter mission \citep{Mueller2020}, one of its remote-sensing instruments, the Extreme Ultraviolet Imager \citep[EUI;][]{Rochus2020}, recorded numerous erupting prominences with its Full Sun Imager (FSI), in the 304~{\AA} and 174~{\AA} channels. A novel aspect of FSI that is particularly important  for studies of eruptive prominences is its uniquely large ($3.8^{\circ} \times 3.8^{\circ}$) field of view (FOV), which allows prominences to be tracked far out in the solar corona.
In this Letter, we report observations of a unique, large prominence eruption that was observed on 15--16 February 2022 and tracked up to heights above $6\,R_{\odot}$ in the 304~{\AA} channel of FSI. This is the first time that an erupting prominence emitting in EUV could be tracked that far in the corona.  


\section{Observations}
\label{S-obs}

\subsection{Kinematics}
\label{S-kinematics}

\begin{figure}[!ht]
\centering
\includegraphics[width=0.49\textwidth]{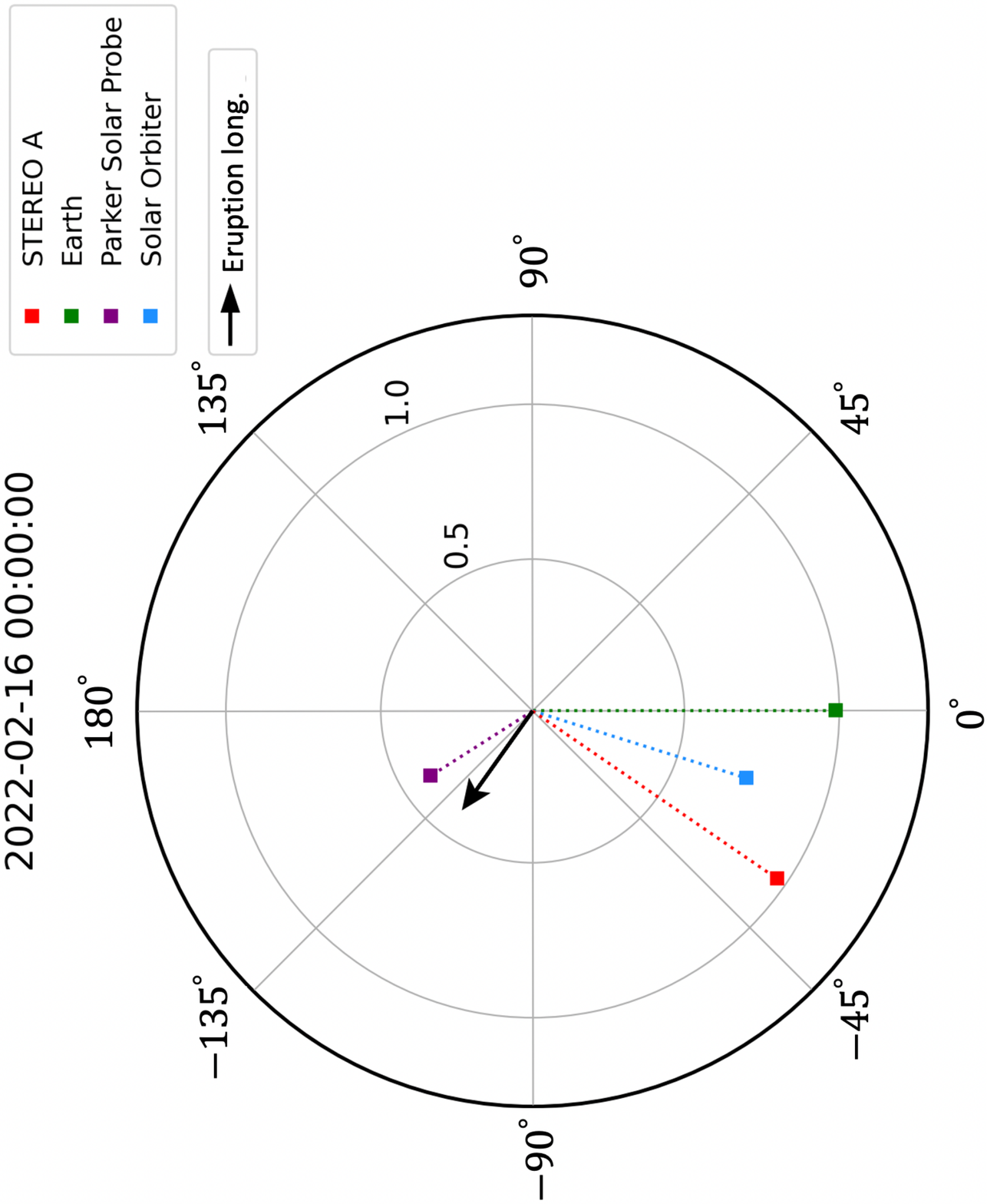}
\caption{Positions of Solar Orbiter, STEREO A, and Parker Solar Probe with respect to Earth on 16 February 2022, 00:00~UT, in heliocentric Earth equatorial  coordinates. The black arrow shows the longitude of the prominence eruption. The image was created using Solar-MACH \mbox{(\url{https://solar-mach.github.io/})}.} \label{F-positionspacecraft}
\end{figure}

\begin{figure*}[p]
\centering
\includegraphics[width=0.83\textwidth]{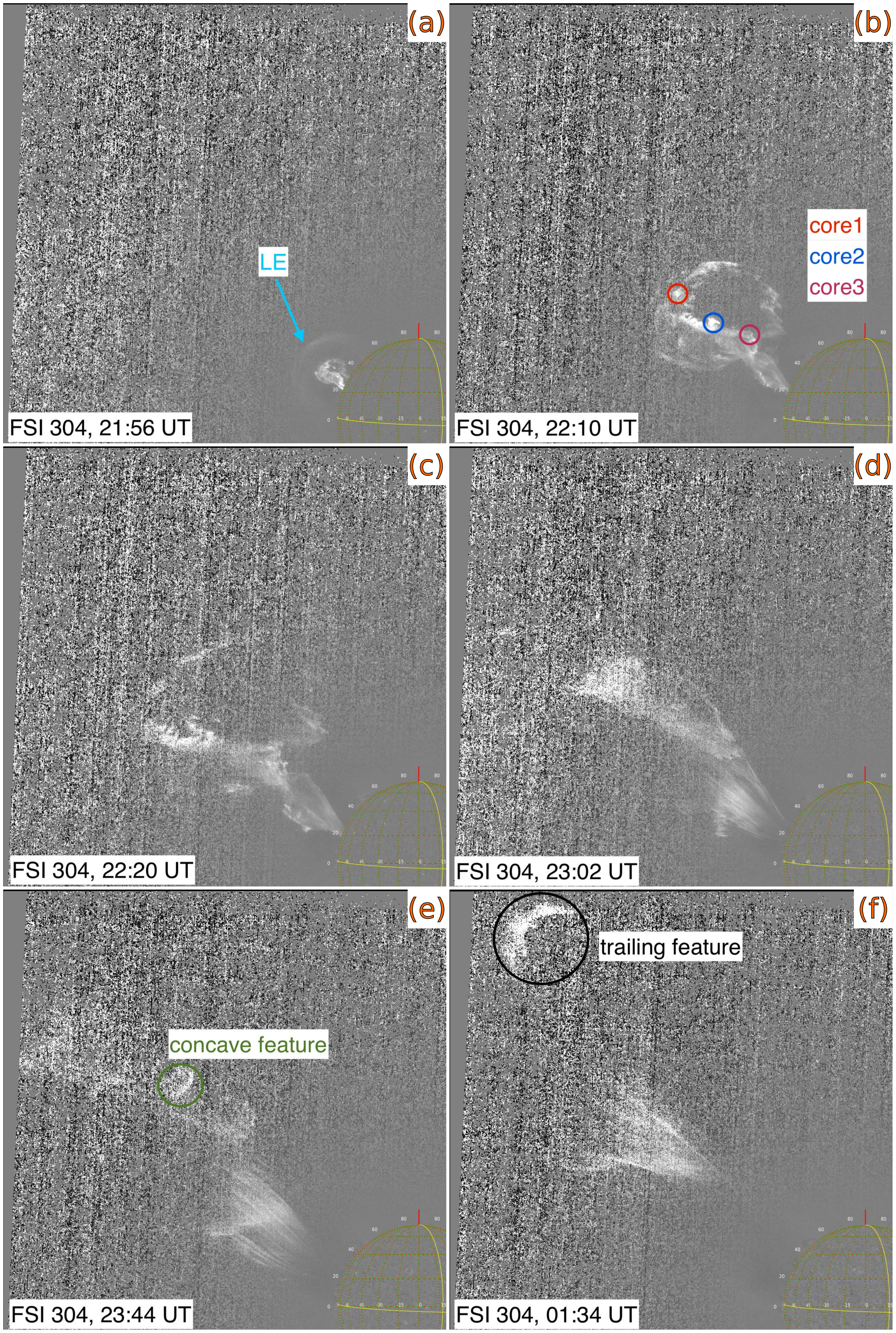}
\caption{Prominence eruption observed by EUI/FSI aboard Solar Orbiter in the 304~{\AA} passband on 15--16 February 2022. The frame taken at 21:50~UT on 15 February 2022 was subtracted from each frame. A few representative features are marked: the LE, three knots in the prominence (core1, core2, and core3), a concave feature, and a trailing feature. These features are used to measure the eruption kinematics shown in Fig.~\ref{F-ht}. The image was created using JHelioviewer \citep{Mueller2017}. An animation of this figure is available online.} \label{F-eruptionfsi304}
\end{figure*}

Positions of the observing spacecraft close to the prominence eruption time (16 February 2022, 00:00~UT) are shown in Fig.~\ref{F-positionspacecraft}. The prominence observed by FSI in the 304~{\AA} channel (FSI304) was a bright and complex event with developed fine structure (Fig.~\ref{F-eruptionfsi304}). The eruption started around 21:50~UT on 15 February, and the prominence was last seen in the FOV of FSI on 16 February around 05:00~UT. The eruption was first seen as a bright prominence with a clear leading edge (LE) ahead of it (Fig.~\ref{F-eruptionfsi304}a), resembling a typical three-part structure of a CME \citep{Illing1985}. Around 01:40~UT on 16 February, a trailing part of the erupting prominence was seen at heights as great as $6.6\,R_{\odot}$ (the outer limit of the FSI FOV; see Fig.~\ref{F-eruptionfsi304}f). A detailed description of the prominence eruption and the associated flare is given in Appendix~\ref{A-detailsobs}. 

\begin{figure*}[!ht]
\centering
\includegraphics[width=0.95\textwidth]{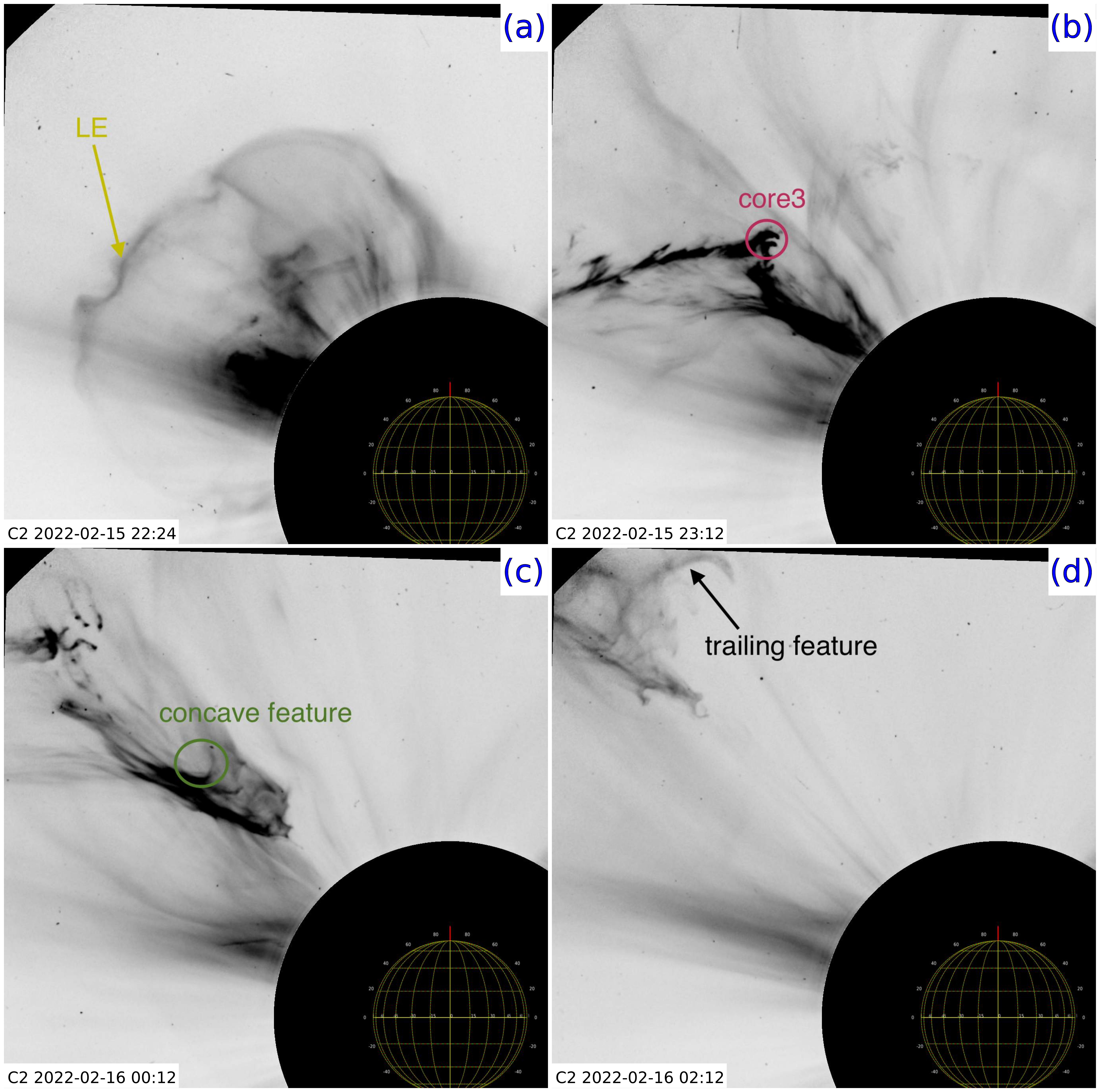}
\caption{Propagation of the associated CME observed by the SOHO/LASCO C2 coronagraph. An inverted colour table is used (bright coronal features appear darker). A few features used for tracking the CME kinematics are marked (cf. Fig.~\ref{F-eruptionfsi304}): the LE of the CME at PA 56$^{\circ}$ (a), the part of the core marked as core3 (b),  a concave feature  (c), and a trailing feature (d). The image was produced using JHelioviewer \citep{Mueller2017}. } \label{F-trackc2}
\end{figure*}

The CME associated with the erupting prominence was observed by the Large-Angle Spectroscopic COronagraph \citep[LASCO;][]{Brueckner1995} aboard the Solar and Heliospheric Observatory \citep[SOHO;][]{Domingo1995} and by the COR2 coronagraph \citep{Howard2008} aboard the Solar--TErrestrial RElations Observatory \citep[STEREO~A;][]{Kaiser2008} Ahead spacecraft. Figure~\ref{F-trackc2} shows that the prominence material constituting the core of the CME could be seen up to ${\sim}7.96\,R_{\odot}$ from the Sun's centre (the edge of the LASCO/C2 FOV). 

\begin{figure*}[!ht]
\centering
\includegraphics[width=0.9\textwidth]{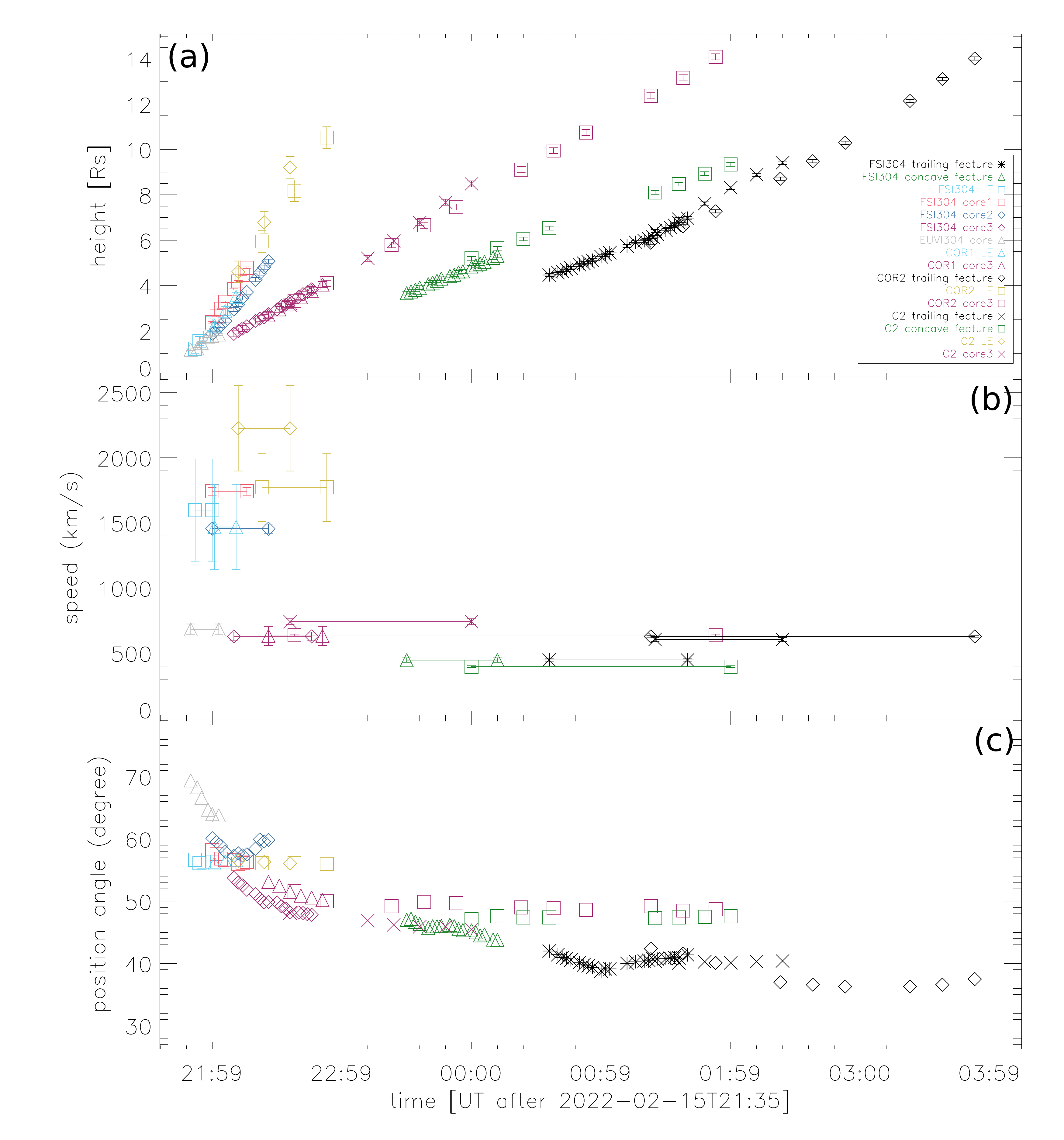}
\caption{Kinematics of various tracked features (listed in Table~\ref{T-trackedfeatures}) associated with the 15--16 February 2022 prominence eruption:  deprojected radial height versus time (a), deprojected radial speeds (b), and evolution of the PA (c).} \label{F-ht}
\end{figure*}

To analyse the eruption kinematics, we tracked several off-limb features (see Figs.~\ref{F-eruptionfsi304} and \ref{F-trackc2}) observed from at least two vantage points to reconstruct their position in three-dimensional (3D) space (Fig.~\ref{F-ht}). The overall tip of the prominence (i.e.\ the CME core) was tracked at a position angle (PA) of ${\sim}56^{\circ}$ in both FSI304 and COR1 data (the PA is measured anti-clockwise from the north pole) and at a PA of ${\sim}65^{\circ}$ in the EUVI304 data.
Three bright knots of the prominence core (core1, core2, and core3) are marked in Fig.~\ref{F-eruptionfsi304}b. The core1 feature is a bright knot at the leading part of the prominence (PA of ${\sim}56^{\circ}$) and was also observed by the Extreme-UltraViolet Imager (EUVI) aboard STEREO~A, the Solar Ultraviolet Imager (SUVI), and the Atmospheric Imaging Assembly (AIA). The core2 feature is a bright knot that could be tracked at the PA of ${\sim}60^{\circ}$. The core3 feature is a kink-like structure that was observed at the PA of ${\sim}50^{\circ}$, also seen by LASCO C2 (PA of ${\sim}46^{\circ}$; see Fig.~\ref{F-trackc2}b), COR1 (PA of ${\sim}51^{\circ}$), and COR2 (PA of ${\sim}49^{\circ}$). The three core features are sufficiently compact for triangulation.

A concave feature was observed after 23:30~UT at a PA of ${\sim}45^{\circ}$. It was also observed by LASCO/C2 at a PA of ${\sim}47^{\circ}$ (see Fig.~\ref{F-trackc2}c).
A trailing feature was tracked at a PA of ${\sim}40^{\circ}$ from 00:36~UT on 16 February onwards. It was also observed by COR2 and C2 (see Fig.~\ref{F-trackc2}d) at approximately the same PA. Besides these features in the CME core, we also tracked the LE of the CME as observed by FSI304 (Fig.~\ref{F-eruptionfsi304}a), LASCO C2 (Fig.~\ref{F-trackc2}a), and STEREO/COR1 and COR2.

The 3D positions were calculated using triangulation, except for the LE, for which we used graduated cylindrical shell (GCS) model reconstruction \citep{Thernisien2009}. The triangulation method requires identification of the same point in the two images \citep[tie-pointing; see][]{Inhester2006}. The 3D positions of lines of sight passing through a point that is visible in the two images are calculated, and the position of the intersection point in 3D space is determined using the scc\_measure.pro program of SolarSoft \citep{Thompson2008}.

\begin{table*}[!ht]
\renewcommand{\arraystretch}{1.2}
\begin{center}
\caption{Features tracked to create the kinematic plots shown in Fig.~\ref{F-ht}.}
\label{T-trackedfeatures}
\begin{tabular}{ccccc}    
  \hline\hline                   
Tracked feature & Description & 3D position & Colour &  Observations\\
  \hline
core1 & bright knot at the prominence front & E151N28 & red & \textit{FSI304}, \textit{EUVI304}, SUVI304, AIA304 \\
core2 & bright knot in the prominence core & E147N27 & blue & \textit{FSI304}, \textit{SUVI304} \\
core3 & kink in the prominence core & E124N40 & magenta & FSI304, \textit{C2}, COR1, \textit{COR2}  \\
core & prominence tip in EUVI304 & E151N28 & grey & \textit{FSI304}, \textit{EUVI304} \\
concave feature & concave feature in the prominence & E128N43 & green & \textit{FSI304}, \textit{C2} \\
trailing feature & seen up to the end of FOV of FSI304 & E125N47 & black & \textit{FSI304}, \textit{C2}, COR2 \\
FSI304 LE & CME leading edge & E132N23 & cyan & \textit{FSI304}, \textit{COR1} \\
C2 LE & CME leading edge & E132N34 & yellow & \textit{C2}, \textit{COR2}  \\
  \hline
\end{tabular}
\end{center}
\tablefoot{
Column 1 identifies the feature, Col. 2 gives a short description, Col. 3 shows the longitude and latitude derived by the triangulation (except for the LEs, which are derived using the GCS reconstruction), Col. 4 shows the colour used to display the feature in the kinematic plots, and Col. 5 lists the instruments that observed the features. The specific pairs of instruments used to perform the triangulation or the GCS reconstruction are marked in italics. 
}
\end{table*}

The GCS model is designed to reproduce the large-scale structure of flux-rope-like CMEs that cannot be determined via triangulation. It consists of a tubular section that forms the main body of the structure and is attached to two cones that correspond to the CME `legs'. Only the surface of the CME is modelled, there is no rendering of its internal structure. This provides information mostly on the propagation of the LE of the CME.

All the tracked features together with their coordinates are listed in Table~\ref{T-trackedfeatures}. We used the calculated longitude and latitude of the features and their measured radial heights seen in projection on the respective planes of the sky of each telescope (not shown) in order to calculate their true heights and velocities. The deprojected height--time plot of the features measured by each telescope is shown in Fig.~\ref{F-ht}a. The errors in height were calculated using Eq.~(1) of \citet{Zhukov2021} and take the finite pixel size of each telescope and the resulting 3D reconstruction error  into account. Figure~\ref{F-ht}a shows a generally good agreement between deprojected heights derived from the projected measurements made by individual telescopes. The remaining discrepancies could be explained by differences in spectral passbands of individual telescopes, and by the lack of the strict simultaneity of the images. The large FOV of EUI/FSI allows the erupting prominence in the 304~{\AA} passband to be tracked to great heights, typically up to above 4~$R_\odot$ (deprojected). The trailing feature in this event is conspicuous, as it is visible in the 304~{\AA} passband of the FSI up to the projected height of $6.64\,R_{\odot}$, corresponding to a true 3D height of $6.97\,R_{\odot}$.

The speed of each feature was calculated by fitting a first-order polynomial to the height--time points.
The deprojected speeds of the tracked features are shown in Fig.~\ref{F-ht}b. The CME LE reaches speeds of ${\sim}2200$~km~s$^{-1}$. The LE is closely followed by the prominence, which accelerates to speeds of ${\sim}1700$~km~s$^{-1}$. The features that erupted earlier (core1 and core2) generally reach speeds that are higher than those of the features erupting later (core3, the concave feature, and the trailing feature). These trailing parts of the prominence are significantly slower, propagating at speeds of ${\sim}500$~km~s$^{-1}$. 

Figure~\ref{F-ht}c shows the evolution of the PA of each tracked feature. The propagation is not fully radial, with the erupting prominence deflected towards the equator, as is often observed \citep{Kilpua2009,Sieyra2020}. The strongest deflections occur at the beginning of the eruption.


\subsection{Radiative properties}
\label{S-signal}

\begin{figure*}[!ht]
\centering
\includegraphics[width=0.9\textwidth]{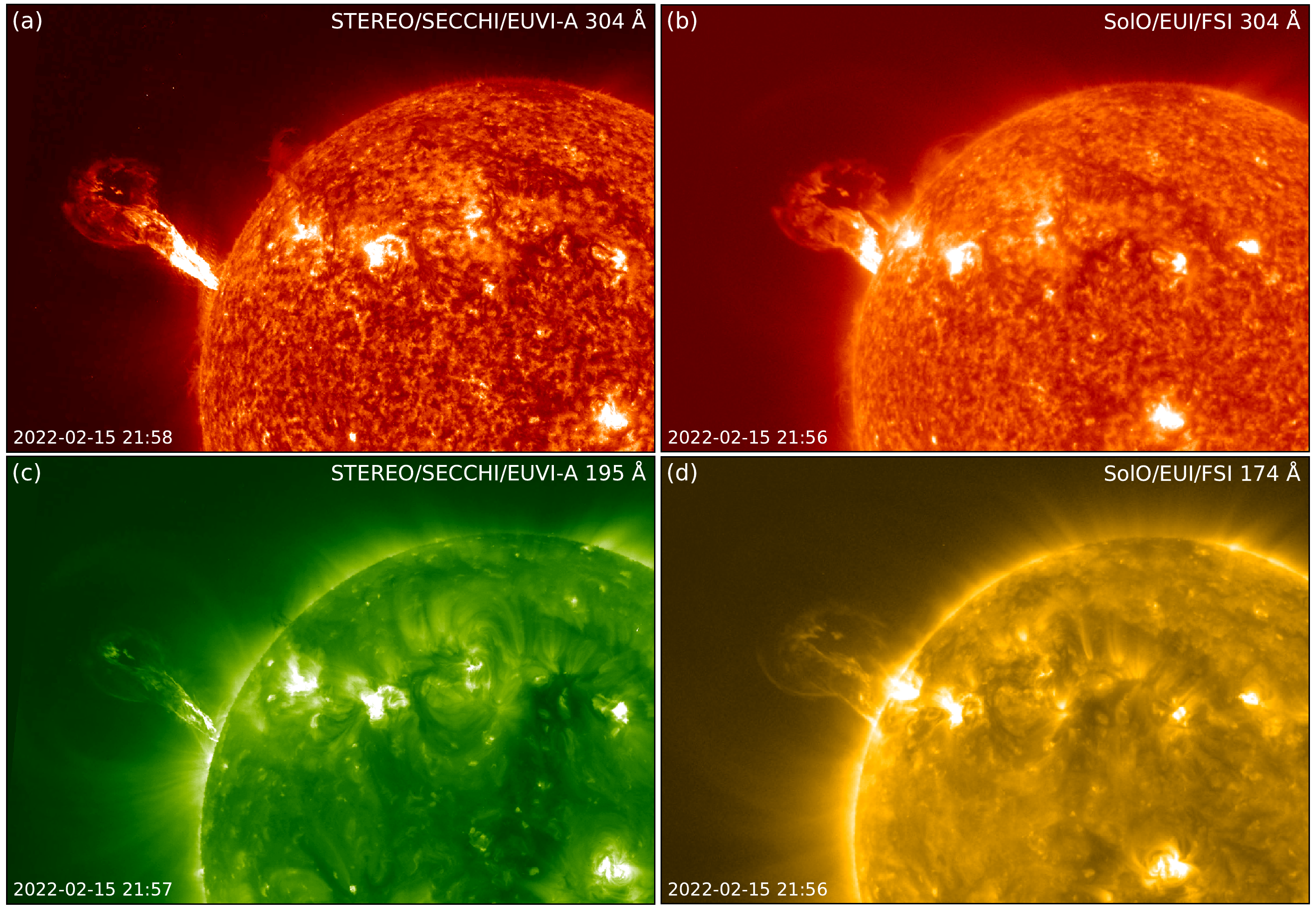}
\caption{Erupting prominence observed on 15 February 2022 in different passbands: STEREO/EUVI 304~{\AA} image (a), Solar Orbiter/EUI/FSI 304~{\AA} image (b), STEREO/EUVI 195~{\AA} image (c), and Solar Orbiter EUI/FSI 174~{\AA} image (d).} \label{F-promin_fsi_euvi}
\end{figure*}

The prominence eruption was also visible in coronal passbands that are usually dominated by the emission of hot (above $\sim$0.5~MK) plasma, for example\ in the 174~{\AA} passband of FSI and the 195~{\AA} passband of EUVI. Figure~\ref{F-promin_fsi_euvi} shows that the prominence was bright (i.e.\ observed in emission), with no sign of absorption that could be seen in the coronal passbands. This may indicate that parts of the eruptive prominence were heated to temperatures above $\sim$0.5~MK and that there is not enough neutral hydrogen and helium to absorb the background coronal radiation. 
The apparent co-location of hot and cool features may be explained by projection effects, as the prominence may have a cooler central part (seen in He~II at 304~{\AA}) surrounded by the hotter PCTR plasma (seen at 174~{\AA}, 195~{\AA}, and in coronal lines present in the 304~{\AA} passband).

In the He~II 304~{\AA} channel, the eruptive prominence is seen clearly by FSI and EUVI (Fig.~\ref{F-promin_fsi_euvi}a--b). The lower part of the prominence is much brighter than the underlying disk, so it is unlikely that the He~II emission there is produced by the resonant scattering of the disk radiation. Additionally, expected dark portions at the upper edge of the prominence \citep[a characteristic feature of the resonant scattering; see][]{Heinzel2015} are not clearly seen. At the observed speeds of around 500~km~s$^{-1}$, the Doppler dimming effect in the 304~{\AA} line will make the resonantly scattered emission negligible \citep{Labrosse2012}.
All this indicates that the excitation of the He~II 304~{\AA} line is probably dominated by collisions rather than by resonant scattering. 

\begin{figure*}[!ht]
\centering
\includegraphics[width=0.9\textwidth]{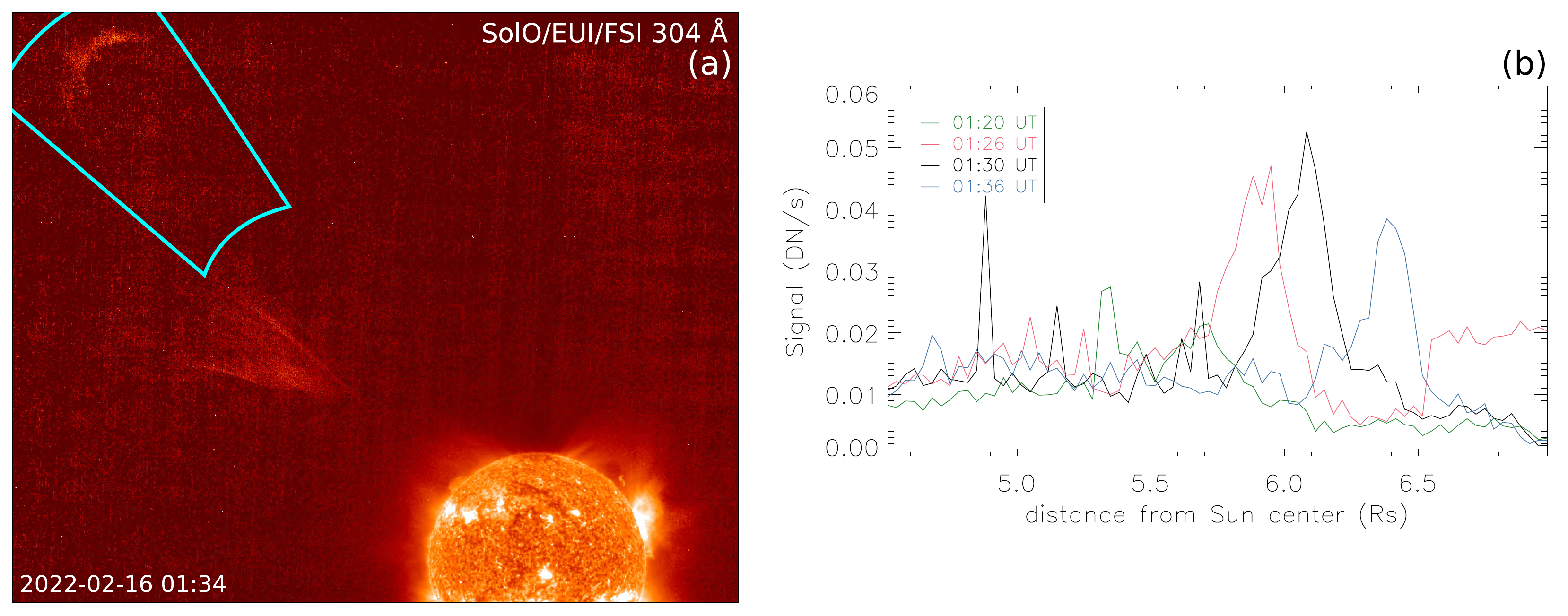}
\caption{Observations of the trailing feature. (a) Solar Orbiter/EUI/FSI image taken in the 304~{\AA} passband at 01:34~UT on 16 February 2022, showing the sector between PAs of 35$^{\circ}$ and 49$^{\circ}$ and radial distances between $4.5\,R_{\odot}$ and $7\,R_{\odot}$, in which the radial brightness profile is calculated. (b) Radial brightness profiles integrated along the azimuthal direction of the sector shown in the left panel, at four different times.} \label{F-prominsignal}
\end{figure*}

An unusual property of this eruption is that the brightness of the trailing feature of the prominence increases with distance, starting from 01:20~UT at the latest, when the feature was at a distance of around $5.72\,R_{\odot}$.  
To demonstrate this, we selected a sector around the trailing feature in the range $4.5$--$7\,R_{\odot}$ and between the PAs 35$^{\circ}$ and 49$^{\circ}$ (Fig.~\ref{F-prominsignal}a). This sector is further divided into azimuthally aligned sub-sectors with a step of $0.033\,R_{\odot}$. The intensity was integrated in each sub-sector, and the resulting curves of intensity versus height are displayed in Fig.~\ref{F-prominsignal}b for four different times. The brightness increases until 01:30~UT, at which point the trailing feature was situated at around $6.1\,R_\odot$, and then decreases. At its peak, the trailing feature of the prominence at 01:30~UT is approximately four times brighter than the background. 

\section{Summary and discussion}
\label{S-discussions}

We have presented observations of a unique prominence eruption that was recorded by the He~II 304~{\AA} channel of the EUI/FSI telescope aboard Solar Orbiter on 15--16 February 2022. Observations from different vantage points allowed us to calculate 3D locations of different prominence features and to derive their deprojected kinematics. Some parts of the prominence could be tracked up to projected heights of $6.64\,R_{\odot}$ (3D heights of $6.97\,R_{\odot}$). This is the first time that a prominence has been observed at such a great height in the FOV of an EUV imager.

The eruption was a far-side event as seen from the available vantage points, so the initiation and acceleration phases were not observed. The prominence velocity reached speeds of ${\sim}1700$~km~s$^{-1}$. While the leading parts of the prominence show a fast propagation, features at the prominence rear end that are seen later on propagate at lower speeds (${\sim}500$~km~s$^{-1}$). This may indicate that magnetic reconnection initiated the fast eruption \citep{Zhang2001}, and later on the prominence material was swept away as a part of the large-scale relaxation of the coronal magnetic field.

The presence of the He~II 304~{\AA} emission at heights above $6\,R_\odot$ indicates that this part of the prominence was not heated to the fully ionised state during its propagation \citep[the corresponding He$^{+}$ ions are only very rarely detected in the solar wind; see][]{Lepri2010}. 
The coronal EUV passbands indicate the presence of hot PCTR plasma, and no absorption signature is observed, probably due to the dominance of the bright PCTR emission. The hydrogen and helium, which are responsible for EUV absorption, are apparently ionised under high-temperature conditions (collisional ionisation). 
The presence of ionised hydrogen and helium is important for the derivation of the prominence mass from visible-light coronagraph observations, for instance by SOHO/LASCO. From coronagraph images, the column electron density can be derived using Thomson scattering calculations, and the prominence column mass along the line of sight can then be deduced, assuming a certain ionisation degree \citep{Kucera2015}.
Moreover, if the helium is mostly ionised at least to the state of He$^{+}$, then there will be almost no neutral helium He~I D$_3$ line emission, which is otherwise typical for prominences \citep{Jejcic2018}. This line is included in the orange channel of SOHO/LASCO C2 \citep{Floyd2019} and in the visible-light channel of Metis \citep{Heinzel2020}, so the relative importance of the spectral line emission in a passband needs to be taken into account in order to derive the prominence density and mass \citep{Mierla2011}. 

The very bright 304~{\AA} prominence emission suggests that the helium is mostly in the He$^{+}$ state, which occurs at temperatures of 0.05--0.08~MK. The observed speeds of around 500~km~s$^{-1}$ are high enough for the Doppler dimming to make the resonantly scattered emission negligible.  We conjecture that the 304~{\AA} line in this eruptive prominence is due to collisional excitation rather than to resonant scattering of incident disk radiation. 
In a future work we will analyse the 304~{\AA} line formation in this unique event using radiometrically calibrated data and relevant non-LTE  modelling (with `LTE' standing for `local thermodynamic equilibrium'). If the resonant scattering is indeed negligible, then it may be possible to address temporal variations of plasma conditions without the bias introduced by Doppler dimming. This can also help in estimating the prominence mass.

The increase in brightness during the outward propagation of the trailing feature of the prominence is difficult to explain without detailed knowledge of the plasma parameters. If the prominence emission is indeed dominated by the collisional mechanism, then one would expect the brightness to increase if the temperature increases (thus producing more He$^{+}$ ions) or if the prominence column density increases due to geometrical rearrangement \citep{Labrosse2012}.



\begin{acknowledgements}
      {Solar Orbiter is a space mission of international collaboration between ESA and NASA, operated by ESA. The EUI instrument was built by CSL, IAS, MPS, MSSL/UCL, PMOD/WRC, ROB, LCF/IO with funding from the Belgian Federal Science Policy Office (BELPSO); the Centre National d’Etudes Spatiales (CNES); the UK Space Agency (UKSA); the Bundesministerium f\"ur Wirtschaft und Energie (BMWi) through the Deutsches Zentrum f\"ur Luft- und Raumfahrt (DLR); and the Swiss Space Office (SSO). We acknowledge the use of PROBA2/SWAP, SDO/AIA, GOES-16/SUVI, SOHO/LASCO, and STEREO EUVI, COR1, COR2 data. I.C.J and J.M. acknowledge funding by the  BRAIN-be project SWiM (Solar Wind Modelling with EUHFORIA for the new heliospheric missions). The ROB team thanks the Belgian Federal Science Policy Office (BELSPO) for the provision of financial support in the framework of the PRODEX Programme of the European Space Agency (ESA) under contract numbers 4000134474, 4000134088, and 4000136424. M.~M. would like to thank J.~Dudik for fruitful discussions on filament eruption mechanisms and S.~Willems for updating the missing data on the local servers. 
      E.~P. acknowledges support from NASA HTMS grant no. 80NSSC20K1274.
      S.~J. acknowledges the support from the Slovenian Research Agency No. P1-0188.
      D.~M.~L. is grateful to the Science Technology and Facilities Council for the award of an Ernest Rutherford Fellowship (ST/R003246/1).
      P.~H. was supported by grant No. 22-34841S of the Czech Funding Agency and acknowledges support from the grant No. 19-16890S of the Czech Science Foundation (GA \v CR).}
\end{acknowledgements}

\bibliographystyle{aa}
\bibliography{bibliography.bib}

\begin{appendix}

\section{Detailed description of observations}
\label{A-detailsobs}

\subsection{Overview of prominence observations}
\label{A-overview}

The positions of different spacecraft close to the eruption time (16 February 2022, 00:00~UT) are given in Table~\ref{T-positionspacecraft} and shown in Fig.~\ref{F-positionspacecraft}. Solar Orbiter was observing from a distance of 0.721~au from the Sun. This means that a solar feature observed by FSI will be observed 2 minutes and 12.6 seconds later by instruments at Earth-orbiting observatories, and 2 minutes and 2.3 seconds later by the instruments aboard the STEREO~A \citep{Kaiser2008} spacecraft. The cadence of the FSI data in the 304~{\AA} and 174~{\AA} passbands was around 2~minutes. The Metis coronagraph \citep{Antonucci2020} aboard Solar Orbiter was not observing at that time. 

\begin{table}[!ht]
\renewcommand{\arraystretch}{1.2}
\begin{center}
\caption{Positions of observing spacecraft on 16 February 2022, 00:00 UT.}
\label{T-positionspacecraft}
\begin{tabular}{cccc}    
  \hline\hline                   
 Vantage & Distance to the Sun & Latitude & Longitude \\
 point & [au] & [deg] & [deg] \\
  \hline
Earth & 0.987 & S07 & E00  \\
Solar Orbiter & 0.721 & S03 & E17  \\
STEREO-A & 0.967 & S04 & E34  \\
PSP & 0.377 & N04 & E146  \\
  \hline
\end{tabular}
\end{center}
\tablefoot{
Column 1 gives the name of the vantage point, Col. 2 the distance to the Sun, Col. 3 the Heliocentric Earth EQuatorial (HEEQ) latitude, and Col. 4 the HEEQ longitude.
}
\end{table}

The 304~{\AA} passband that is mainly used in our study is dominated by the He~II Ly-$\alpha$ emission line at 303.78~{\AA} formed by the He$^{+}$ ions. This ion is populated by photoionisation of the neutral helium at lower temperatures, but at higher PCTR temperatures the collisional ionisation will dominate. Then, having enough populated He$^{+}$, the 304~{\AA} line is formed by the resonance scattering of the disk radiation in a low-temperature and low-density regime, and will be enhanced at higher PCTR temperatures. However, at high speeds of a few hundred km~s$^{-1}$, the Doppler dimming effect will substantially lower the scattering component in the 304~{\AA} line source function.
The passband also contains a number of weaker coronal lines, most prominently the Si~XI line at 303.33~{\AA} formed at temperatures of around 2~MK \citep{Delaboudiniere1999, Labrosse2012}. Although any quantitative separation of coronal and prominence emission is difficult without spectroscopic observations, hot (coronal) and cool (prominence) structures can be qualitatively distinguished in the images based on morphology: smooth for coronal features, ragged for prominence features. We also see the prominence-like ragged structures (corresponding to the PCTR) in other hot coronal channels, such as 195~{\AA} or 171~{\AA} (Fig.~\ref{F-promin_fsi_euvi}; see \citealt{Parenti2012}). Resonance continua of neutral hydrogen and helium are probably too weak to contribute to the 304~{\AA} channel.

Other EUV telescopes that observed the eruption were: the EUVI \citep{Howard2008} aboard the STEREO~A spacecraft, with a cadence of ${\sim}3$~min; the AIA \citep{Lemen2012} aboard the Solar Dynamics Observatory \citep[SDO;][]{Pesnell2012}, with a cadence of 12~s; the SUVI \citep{Darnel2022} aboard Geostationary Operational Environmental Satellite-16 (GOES-16), with a cadence of ${\sim}4$~min; and the Sun Watcher using Active Pixel \citep[SWAP;][]{Seaton2013} sensor aboard the PRoject for On Board Autonomy 2 \citep[PROBA2;][]{Santandrea2013} spacecraft, with a cadence of ${\sim}2$~min. The corresponding CME was subsequently observed in white light by the COR1 and COR2 coronagraphs \citep{Howard2008} aboard STEREO~A, with cadences of 5~min and 15~min, respectively, and by LASCO \citep{Brueckner1995} aboard SOHO \citep{Domingo1995}, with a 12-minute cadence. 

From triangulation (see Sect.~\ref{S-kinematics}), the positions of different prominence features span a longitudinal range from approximately E125 to E150 and a latitudinal range from around N20 to N50 (Stonyhurst coordinates). This means that it was a far-side event for Solar Orbiter, Earth, and STEREO~A, and it erupted in the direction of the Parker Solar Probe. After the eruption, a system of bright post-eruption loops (a flare) appears, seen better from the STEREO~A perspective. A short description of the source location and the associated flare is presented in Sect.~\ref{A-flare}. The prominence structure as seen by FSI304 is clearly recognisable in the LASCO C2 images, as the angular separation between Solar Orbiter and SOHO is only 17$^{\circ}$. 

\subsection{Observations from the Solar Orbiter perspective}
\label{A-soloobs}

\begin{figure*}[!th]
\centering
\includegraphics[width=1.0\textwidth]{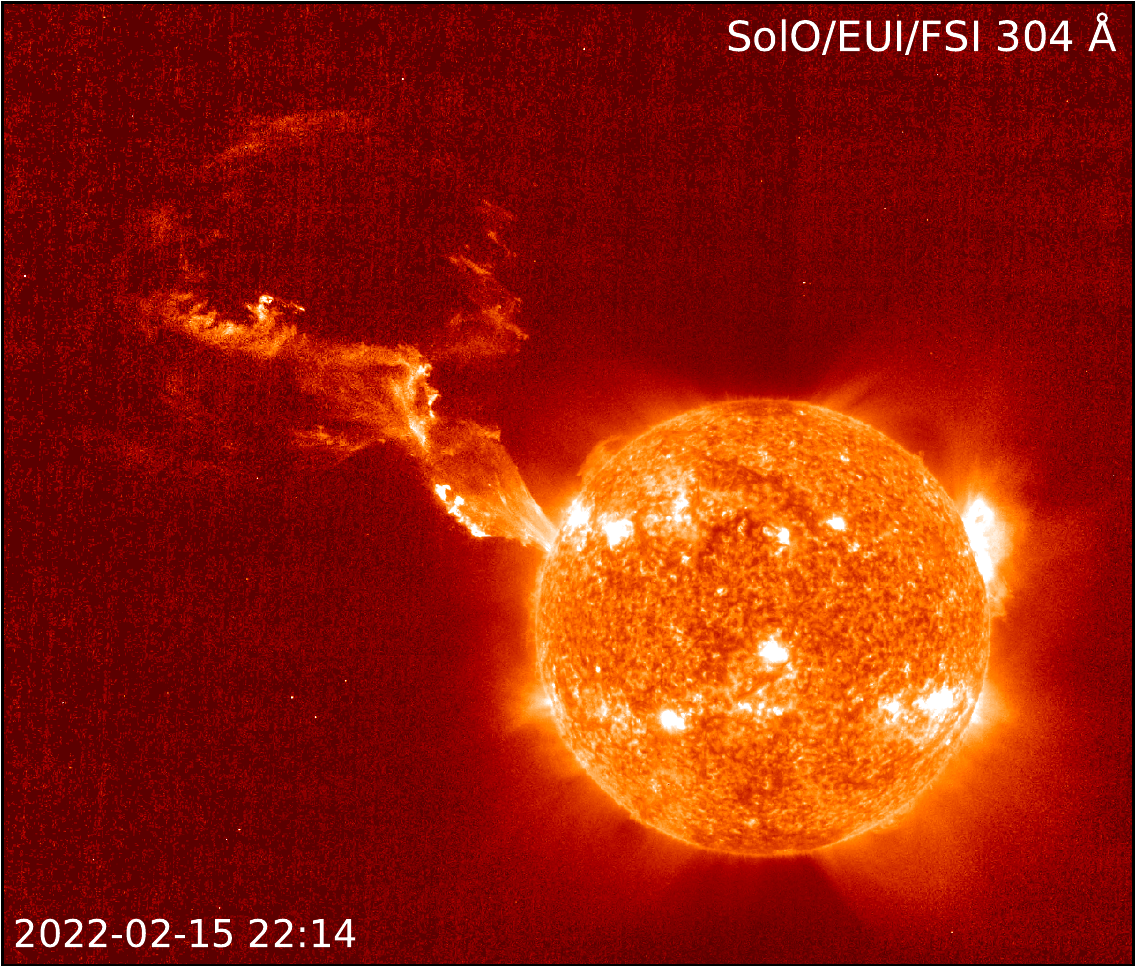}
\caption{Erupting prominence observed by EUI/FSI aboard Solar Orbiter in the He~II 304{~\AA} passband on 15~February 2022, at 22:14~UT. The off-limb emission has been enhanced with a radial filter. An animation of this figure is available online.} \label{F-fsi304enhanced}
\end{figure*}

The prominence eruption was observed by FSI304 up to heights as great as $6.64\,R_{\odot}$ (see Fig.~\ref{F-fsi304enhanced} and the corresponding animation). The eruption was also seen by FSI in its 174~{\AA} channel (FSI174; see Fig.~\ref{F-promin_fsi_euvi}d). The morphology of the eruption was similar (consisting of the LE and the core), although not so clearly visible as the 174~{\AA} passband is mostly sensitive to the emission of the hotter plasma at temperatures of around 1~MK. According to the FSI174 observations, it was visible only up to heights of around $3.35\,R_{\odot}$ (at 23:26~UT). In the FSI304 image, one leg was connecting the leading part of the prominence to the Sun (Fig.~\ref{F-eruptionfsi304}b). Flows along the leg were observed in different structures for quite some time after the leading part was last visible in the FSI FOV (${\sim}$22:20~UT). The leg seemed to disconnect from the Sun at ${\sim}$23:30~UT.

\subsection{Observations from the Earth perspective}
\label{A-earthobs}

The eruption was observed in all SDO/AIA channels, except for 94~{\AA}, where it was not visible at all, and 335~{\AA}, where only a faint off-limb LE was visible. 
Off-limb dimming and a large on-disk `EIT wave'\footnote{`EIT' stands for the Extreme-ultraviolet Imaging Telescope, see \citet{Delaboudiniere1995}.} or EUV wave \citep[see e.g. reviews by][]{Zhukov2011, Liu2014} were observed by AIA at 193~{\AA}. 
The eruption was clearly seen in the 304~{\AA}, 171~{\AA}, 131~{\AA}, and 211~{\AA} channels. The eruption was seen up to a height of ${\sim}1.48\,R_{\odot}$, around the edge of the AIA FOV. 

PROBA2/SWAP observed the eruption up to a height of ${\sim}1.87\,R_{\odot}$. Higher up, the weak prominence emission could no longer be distinguished against the noisy background.
The eruption was also observed by GOES-16/SUVI in the 304~{\AA} channel up to $1.88\,R_{\odot}$ (the limit of the SUVI FOV). 

The CME associated with the erupting prominence was observed as a full halo from the SOHO/LASCO vantage point. Figure~\ref{F-trackc2} shows that the prominence material constituting the core of the CME could be seen up to ${\sim}7.96\,R_{\odot}$ from the Sun's centre (the edge of the LASCO C2 FOV). The prominence features were observed as high as $30.80\,R_{\odot}$ in the FOV of the LASCO C3 coronagraph (not shown). Furthermore, LASCO also observed signatures of a shock wave propagating ahead of the LE of the CME.

\subsection{Observations from the STEREO A perspective}
\label{A-staobs}

The EUVI telescope aboard STEREO~A observed the eruption in all four of its channels (304~{\AA}, 171~{\AA}, 195~{\AA}, and 284~{\AA}) up to ${\sim}1.73\,R_{\odot}$, at which point the prominence could no longer be distinguished against the background. 
A dimming and an EIT wave (EUV wave) were observed off the limb and on the disk in the 195~{\AA} and 284~{\AA} channels.
An unwinding (in the clockwise direction as seen from above) can be observed at the base of the erupting prominence shortly after the start of the eruption. 
The EUVI images used for triangulation were binned to 1024$\times$1024 pixels in order to have approximately the same spatial resolution as the FSI304 images.
The prominence constituting the core of the associated CME was observed by the COR1 and COR2 coronagraphs up to a height of ${\sim}15\,R_{\odot}$.


\subsection{Observations of the associated flare}
\label{A-flare}

\begin{figure*}[!th]
\centering
\includegraphics[width=0.89\textwidth]{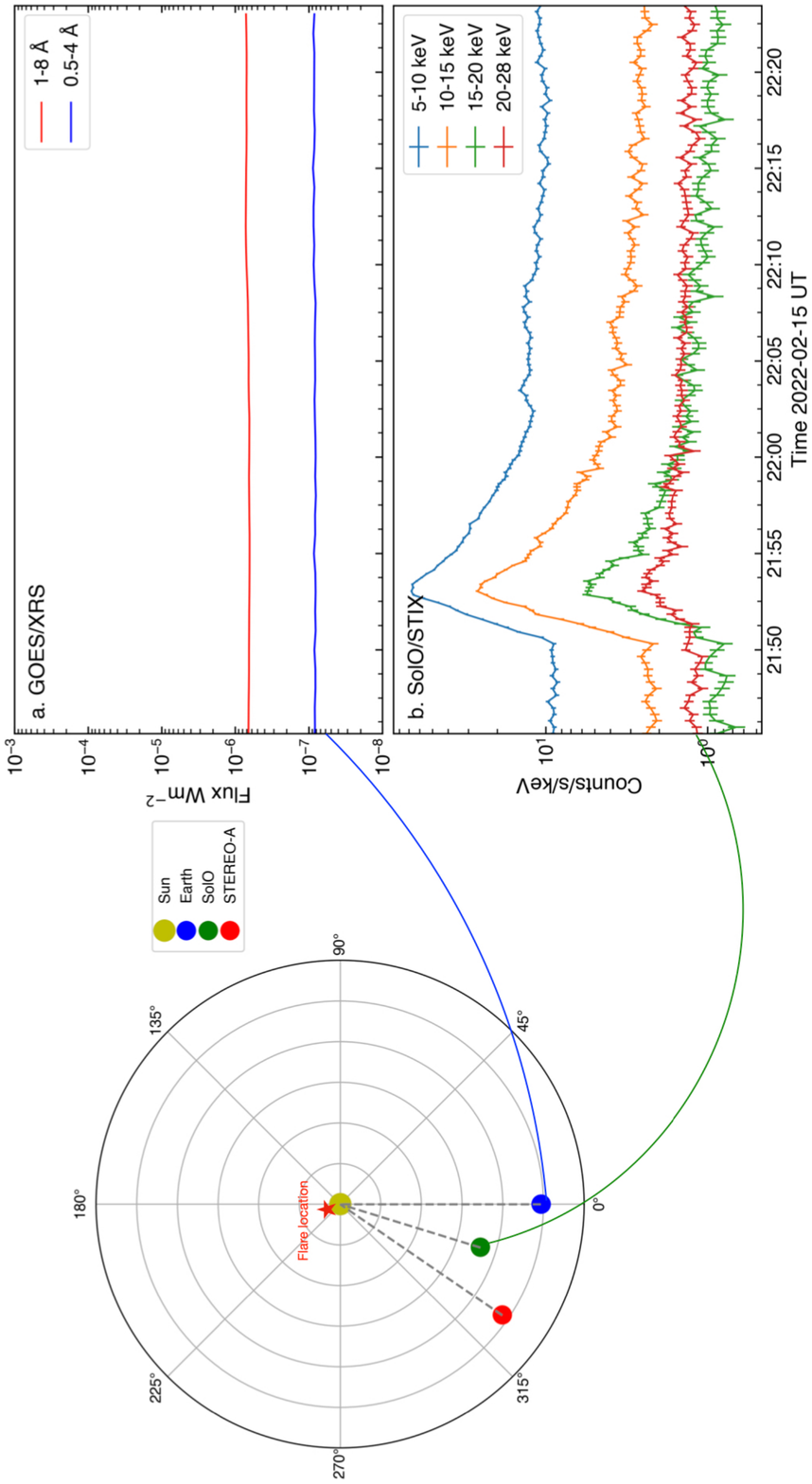}
\caption{X-ray observations of the flare associated with the prominence eruption on 15--16 February 2022 from different vantage points. The relative positions of Solar Orbiter and Earth together with the estimated location of the flare associated with the eruption are shown on the left. The X-ray emission measured by XRS aboard the Earth-orbiting GOES spacecraft is shown in the top-right plot. No enhancement in emission is seen, confirming that the flare was completely occulted as seen from Earth. The Solar Orbiter/STIX observations in four different X-ray bands are shown in the bottom-right plot. Given the position of Solar Orbiter, STIX has a better view of the flare above the limb, and clear X-ray signatures can be seen at energies up to 28~keV.} \label{F-stix}
\end{figure*}

There are no direct observations of the source region of this far-side eruption. Helioseismograms\footnote{\url{http://jsoc.stanford.edu/data/timed/index.html}} (see \citealt{Zhao2019}), based on the data taken by the Helioseismic and Magnetic Imager \citep{Schou2012} aboard SDO, indicate a substantial active region present on 15--16 February 2022 about 60$^{\circ}$ behind the east limb as seen by SDO. This region corresponds to the location of the old sunspot group complex NOAA AR 2936--2938. NOAA 2936 was a rather big sunspot region that showed considerable flux emergence while rounding the north-west solar limb on 5 February.

A post-eruption system of loops is observed by EUVI in the 195~{\AA} passband at the limb starting at ${\sim}$23:30~UT on 15 February. The flare is occulted, and hence no X-ray emission is seen from either of the GOES X-ray Sensor (XRS) channels, as shown in Fig.~\ref{F-stix}. From the perspective of Solar Orbiter, the X-ray emission was observed by the Spectrometer/Telescope for Imaging X-rays \citep[STIX;][]{krucker2020}, as shown in Fig.~\ref{F-stix}. The footpoints of the flare associated with this eruption are most likely occulted by the limb as seen from STIX. This means that the observed emission is most likely thermal soft X-ray emission from the corona (perhaps with a small non-thermal contribution), either from the flare arcade or from the hottest part of the erupting CME. As the majority of the X-ray emission is occulted, it is not possible to conclusively determine the estimated GOES class from these observations. Further detailed analysis of this event as observed by STIX will be presented in a separate paper.

\end{appendix}

\end{document}